\documentclass[12pt,letterpaper]{JHEP3}         
\usepackage{epsfig}

\newcommand{\bea}{\begin{eqnarray}}
\newcommand{\eea}{\end{eqnarray}}
\def\be{\begin{eqnarray}}
\def\ee{\end{eqnarray}}
\newcommand{\nn}{\nonumber}
\newcommand\para{\paragraph{}}

\newcommand{\eqn}[1]{(\ref{#1})}

\newcommand\half{{\ensuremath{\frac{1}{2}}}}

\def\le{\left}
\def\ri{\right}

\newcommand\ov{\over }

\def\Dslash{\,\,{\raise.15ex\hbox{/}\mkern-12mu D}}
\def\Dbarslash{\,\,{\raise.15ex\hbox{/}\mkern-12mu {\bar D}}}
\def\delslash{\,\,{\raise.15ex\hbox{/}\mkern-9mu \partial}}
\def\delbarslash{\,\,{\raise.15ex\hbox{/}\mkern-9mu {\bar\partial}}}
\def\pslash{\,\,{\raise.15ex\hbox{/}\mkern-9mu p}}
\def\calDslash{\,\,{\raise.15ex\hbox{/}\mkern-12mu {\cal D}}}

\newcommand{\RN}{Reissner-Nordstr\"om}

\def\lae{\mathrel{\mathop{\smash{\lower .5 ex \hbox{$\stackrel<\sim$}}}}}
\def\lae{\mathrel{\mathop{\smash{\lower .5 ex \hbox{$\stackrel>\sim$}}}}}

\preprint{CERN-PH-TH/2013-233, DAMTP-2013-56}

\title{Holographic Lattices Give the Graviton a Mass}

\author{Mike Blake${}^1$, David Tong${}^1$ and David Vegh${}^2$\\

${}^1$ Department of Applied Mathematics and Theoretical Physics,
 University of Cambridge, UK
 \\ ${}^2$ Theory Group, Physics Department, CERN, CH-1211 Geneva 23, Switzerland
\\ {\ } \\
{\tt m.a.blake, d.tong@damtp.cam.ac.uk, david.vegh@cern.ch}
}

\abstract{We discuss the DC conductivity of holographic theories with translational invariance broken by a background lattice. We show that the presence of the lattice induces an effective mass for the graviton via a gravitational version of the Higgs mechanism. This allows us to obtain, at leading order in the lattice strength, an analytic expression for the DC conductivity in terms of the size of the lattice at the horizon. In locally critical theories this leads to a power law resistivity that is in agreement with an earlier field theory analysis of Hartnoll and Hofman.}

\begin{document}
\pdfoutput=1
\pagestyle{plain} \setcounter{page}{1}
\newcounter{bean}
\baselineskip16pt

\section{Introduction}

Here's a simple question. Take a quantum field theory, heat it up and throw in a background density of charged stuff. If you pass a DC current through this system, what is the resistance?

\para
If the field theory has translational invariance, this simple question has a simple answer: the resistance is zero and the material is a perfect conductor. This is true for trivial reasons. Translational invariance implies momentum conservation which, in turn, means that there is no mechanism to dissipate the current. To extract something more interesting, we have to work a little harder and introduce effects that break the translational symmetry such as impurities or a background lattice.

\subsubsection*{Umklapp Processes}

Progress can be made if the breaking of translational invariance does not change the infra-red fixed point of the theory. This means that, from the IR perspective, the effects can be captured by the addition of irrelevant operators ${\cal O}$ to the Hamiltonian, which is schematically of the form
\be H = H_0+ \epsilon \,{\cal O}(k_L)\nn\ee
where $k_L$ is the characteristic momentum of the underlying lattice or impurity.
It was shown in \cite{nernst,sandiego} that such an interaction gives rise to  momentum relaxation rate, $\Gamma$, and hence resistivity, given by the retarded Green's function,
\be \Gamma \sim \epsilon^2k_L^2 \lim_{\omega \rightarrow 0} \frac{{\rm Im}\,G^R_{{\cal O}{\cal O}}(\omega,k_L)}{\omega}\label{sd}\ee
This is an interesting formula. Because it involves the spectral density of the operator ${\cal O}$ at momentum $k_L$, if
there is to be any significant momentum dissipation --- say, enough to give the resistivity $\rho$ a power-law dependence on temperature $T$ ---
then there must be low-energy $\omega \rightarrow 0$ excitations at momentum $k_L$.  If not, the relaxation rate
will be Boltzmann suppressed.

\para
Fermi surfaces provide a natural context in which one has light degrees of freedom at finite momentum.  Such modes are simply electrons scattering across the Fermi surface with a net momentum transfer.  Applying \eqn{sd}, with the operator ${\cal O}$ taken to be the four-fermion Umklapp operator, reproduces the well known $\rho \sim T^2$ behaviour of the resistivity of Fermi liquid theory.

\para
There is another, more exotic, way to get low-energy modes at finite momentum. At critical points, excitations have a typical dispersion relations $\omega \sim k^z$, with $z$ the dynamical exponent. In the limit $z\rightarrow \infty$, this dispersion relation broadens out. Such theories are known as {\it locally critical} and arise naturally in the framework of holography in the guise of infra-red AdS${}_2$ regions of spacetime. In such theories, time scales but space does not and the dimension of an operator ${\cal O}(k_L)$ is dependent on the momentum $k_L$. In \cite{sandiego}, Hartnoll and Hofman showed that, when applied to such local critical theories, the formula \eqn{sd} gives a power-law resistivity
\be \rho \sim T^{2\Delta_{k_L}}\label{tscaling}\ee
where the exponent, $\Delta_{k_L}$ is the frequency space scaling dimension of the operator and depends on the lattice spacing $k_L$. 

\para
The arguments of \cite{sandiego} sketched above are purely field theoretic. Given that locally critical theories arise naturally in holography, one can also try to derive the scaling \eqn{tscaling} using holographic methods alone. The appropriate holographic lattices were introduced in \cite{jorge1}  where Einstein's equations were solved numerically (see also \cite{jorge2,jorge3,dil} for related work). Here strong evidence was presented that the DC conductivity indeed obeys \eqn{tscaling} with ${\cal O}$ given by the charge density. However, this evidence relied heavily on numerics. The purpose of the present paper is, in part, to gain an analytic understanding of this scaling behaviour in a purely holographic framework.  Before describing this, there is another thread that we would like to weave into the discussion.

\subsubsection*{Massive Gravity}

A different approach to incorporating momentum dissipation in holographic models was introduced in \cite{vegh}. The basic idea is straightforward: momentum conservation in the boundary theory follows from diffeomorphism invariance in the bulk. If you want to model a theory without momentum conservation, you need to consider a bulk theory without diffeomorphism invariance. Such theories usually go by the name of  {\it massive gravity}.

\para
The closet of massive gravity contains both skeletons and ghosts. There has been recent progress in constructing a (seemingly) consistent theory of a propagating massive spin 2 particle \cite{derahm2}. However, in the context of holographic massive gravity, life is likely to be somewhat easier. To capture momentum dissipation, you only need to give a mass to the gravitons with polarisation parallel to the boundary. This means that the bulk theory retains diffeomorphism invariance in both time and radial directions. In particular, since the timelike components of the graviton do not get a mass, it seems likely that the constraints imposed by ghosts are much weaker, if not completely absent.

\para
The appeal of massive gravity is that, in contrast to explicit lattices or impurities, it is analytically tractable. Moreover, various aspects of thermodynamics and transport in holographic massive gravity have been explored and give encouragingly sensible answers. The low-frequency optical conductivity exhibits a Drude peak \cite{vegh,davison}, with the momentum relaxation rate of the boundary theory determined by the graviton mass \cite{davison,us}. In particular, a universal formula for the DC conductivity was presented in \cite{us}. This formula, which holds at finite temperature and chemical potential, relates the resistivity of the boundary field theory to the mass of the graviton evaluated on the horizon of the bulk black hole. 

\para
Massive gravity provides a phenomenological way to implement momentum dissipation in holography. But its microscopic origins remain mysterious and it is unclear how one can derive it from better motivated models. A second goal of this paper is to shed some light on this.

\subsubsection*{Synthesis}

The purpose of this short note is to draw these threads together. We start by considering Einstein-Maxwell theory in AdS${}_4$, coupled to a neutral scalar field. Translational invariance is  broken by introducing a spatially modulated source for the scalar; this is precisely the set-up studied in  \cite{jorge1}. However, rather than solving the bulk equations numerically, we instead work perturbatively in the strength of the background lattice. We will see that, to leading order, the bulk conductivity calculation simplifies tremendously, with only a handful of fields responding to an applied electric field on the boundary.  

\para
Foremost among the bulk modes is a phonon --- a Goldstone boson arising from the lattice.
Because of bulk diffeomorphism invariance, this phonon is eaten by the metric to give an extra propagating graviton degree of freedom. The net result is a Higgs mechanism for gravity, with the graviton gaining a radially-dependent effective mass, determined by the profile of the bulk lattice. We will show that the equations describing the perturbations of the holographic lattice coincide with those arising from massive gravity. This allows us to import the result of \cite{us}, relating the resistivity to the mass of the graviton at the black hole horizon. Our punchline is that this formula reproduces the expected temperature dependence that arises from \eqn{tscaling} in locally critical theories.

\section{The Holographic Lattice}

We work with the familiar Einstein-Maxwell theory in $d=3+1$ dimensions, with negative cosmological constant. We add to this a neutral scalar field, $\phi$.
\be S_1 = \int d^4x\sqrt{-g}\  \left[\frac{1}{2\kappa^2}\left(R
+\frac{6}{L^2}\right)-\frac{1}{4e^2}F_{\mu\nu}F^{\mu\nu} -\frac{1}{2}g^{\mu\nu}\partial_\mu\phi\,\partial_\nu \phi - \frac{m^2}{2}\phi^2  \right]\nn\ee
We will choose $m^2\leq 0$ so this field corresponds to a relevant or marginal operator, ${\cal O}$, in the boundary theory.

\para
The workhorse solution for applications of holography is the \RN \ black hole, describing the boundary field theory at temperature $T$ and chemical potential $\mu$. This will be our starting point. When $T\ll \mu$, it is well known that  the  solution asymptotes to an AdS${}_2\times {\bf R}^2$ geometry in the infra-red.  This reflects the fact that the boundary theory flows to a  locally critical fixed point.

\para
We now break translational invariance by introducing a spatially modulated source for the operator ${\cal O}$. For static solutions, $\phi_0(x,y,r)$, the near the boundary expansion of the scalar wave-equation takes the form
\be
\phi_0(r,x,y) \sim \phi_{-}(x,y) \bigg(\frac{r}{L}\bigg )^{\Delta_{-}} + \,\phi_{+}(x,y) \bigg(\frac{r}{L} \bigg) ^{\Delta_{+}}+\ldots
\ee
where $\Delta_{\pm} = \frac{3}{2} \pm \sqrt{\frac{9}{4} + m^2 L^2}$. For technical simplicity, we will  work with the standard quantisation which means that we impose a  source by fixing the leading fall-off, $\phi_{-}$. Here we choose to work with the striped source
\be
\phi_{-} = \epsilon  \cos(k_L x) \label{stripe}
\ee
where $\epsilon$ is a small number that allows us to treat the lattice perturbatively. Turning on this source is equivalent to turning on a spatially modulated potential in the boundary theory,  somewhat analogous to the optical lattices in cold atom experiments. As usual the subleading fall-off, $\phi_{+}$, has the interpretation of the expectation value of the dual operator $\cal O$ in the boundary theory. 
\para 
The radial profile of the lattice is dynamically determined by the scalar wave equation in the bulk. At leading order in $\epsilon$, we can work with the \RN \ geometry. The bulk solution takes the form $\phi(r,x,y) = \epsilon\,\phi_0(r)\cos(k_L x)$, where the background lattice profile $\phi_0(r)$ satisfies 
\be
\frac{d}{dr}\left(\frac{f}{r^2} \frac{d\phi_0}{dr}\right) - \frac{k_L^2}{r^2} \phi_0 -\frac{m^2 L^2}{r^4}\phi_0 = 0  \label{background}
\ee
with $f(r)$ the familiar emblackening factor of the \RN \ metric\footnote{For those who have forgotten, $f(r) = 1 - r^3/r_h^3 -  \mu^2 r^3 / 4r_h + {\mu^2 r^4}/{4r_h^2}$. See, for example, \cite{sean,john} for a review of this background and some of its many applications.}. 

\para
 Because the operator ${\cal O}$ is relevant (or possibly marginal) one might expect that, once sourced, the profile $\phi_0$ will grow in the infra-red. Indeed, as one moves away from the boundary, $\phi_0$ does  begin to grow. However, the homogeneous mode of ${\cal O}$ is not sourced by \eqn{stripe} and this changes the expected behaviour of the perturbation under RG flow. As one moves yet further into the bulk, the finite wave-vector corrections become important and $\phi_0$ develops a maximum. By the time one reaches the infra-red AdS${}_2\times {\bf R}^2$ geometry, the scalar field is decaying: it is dual to an irrelevant operator in the locally critical theory\footnote{Actually, for $m^2 < 0$, it is necessary for $k_L$ to be sufficiently large in order for $\phi$ to be irrelevant in the IR. This is the case we consider in this note.}.

\para 
This behaviour means that the scalar field $\phi$ is bounded everywhere in the bulk, with its size controlled by $\epsilon$. This makes  a perturbative treatment possible. Our goal in this paper is to calculate the resistivity due to the lattice to order $\cal{O}$$(\epsilon^2)$.  By turning on a lattice in the scalar field, as opposed to the chemical potential, we have ensured that the stress tensor of our lattice is smaller than the lattice itself - that is $\cal{O}$$(\epsilon^2)$. As we now explain, the key benefit of this is that it allows us to neglect the back reaction of the lattice on the background geometry. 
\para In principle, the metric and the gauge field will receive corrections due to the back reaction that can be expanded as power series in $\epsilon$. To compute the conductivity to ${\cal O}(\epsilon^2)$, we must expand the metric and gauge field as
\bea
 & g_{\mu \nu} =   & g_{\mu \nu}^0(r)    +  \epsilon^2 \le[ {g_{\mu \nu}^H}(r) + {g_{\mu \nu}^{I}}(r) \cos(2 k_L x) \ri] + \ldots \nn  \\
 &  A_{\mu} =  & A_{\mu}^0(r)  + \epsilon^2 \le[ A_{\mu}^{H}(r)+ A_{\mu}^{I}(r) \cos (2 k_L x) \ri] + \ldots   \nn
\eea
The corrections contain both homogeneous (e.g $A_{\mu}^H$) and inhomogeneous (e.g $A_{\mu}^I$) pieces.  For our purposes,  the inhomogeneous components can  be neglected because they can only contribute to the zero-momentum conductivity equations after interacting with  another oscillation, after which they become $\cal{O}$$({\epsilon^4})$. In contrast, the homogenous parts of the background do enter the conductivity equations at $\cal{O}$$(\epsilon^2)$. Nevertheless, it was shown in \cite{us}  that, in the context of massive gravity, the DC conductivity is independent of the corrections to the background. We will see shortly that the same result holds here too. This means that, to leading order, the DC conductivity is blind to all these corrections to the background.

\para The net result of these simplifications is that, in order to compute the DC conductivity,  we may treat the background geometry as being the \RN\ black hole, with an oscillating scalar lattice sitting on top\footnote{There is, in fact, an even simpler system in which this is the true solution: this is a complex scalar field with $\psi \sim e^{ik_Lx}$ source so that the modulation cancels out in the stress tensor. A related background was discussed in \cite{metal}.}.

%
%
%
%
%

\subsubsection*{Shake it}

We now perturb the lattice background to determine the conductivity. We do this by adding a small electric field on the boundary of the form $\delta A_x e^{-i\omega t}$. We impose ingoing boundary conditions at the IR horizon to determine the solution  $\delta A_x(r;\omega)$, the optical conductivity is then given by
\be
\sigma(\omega) = \left.\frac{1}{e^2}\frac{\delta A_x'}{i \omega \delta A_x} \right|_{r=0}\label{cond}
\ee
To compute the DC conductivity, we must work at finite $\omega$ and, at the end, take the limit $\omega\rightarrow 0$.  Importantly, however, all perturbations have zero momentum. This simple fact  will help us below.

\para
In the usual case of a homogeneous black hole, $\delta A_x$ sources a metric perturbation $\delta g_{tx}$  but, if we work in gauge $\delta g_{rx}=0$,  nothing more. In contrast, in the full lattice background, studied numerically in \cite{jorge1}, things are much more complicated and one ends up having to solve for 11 coupled perturbations. Thankfully, in our small-lattice expansion, things are not so bad. We can continue to work in the gauge $\delta g_{rx}=0$. We have already argued that to leading order it is consistent to treat the background metric and gauge field as homogeneous. As a result, the metric perturbation $\delta g_{tx}$ sources an inhomogeneous scalar perturbation $\delta \phi$, but there things stop\footnote{For example: the scalar perturbation $\delta\phi$ sources an inhomogeneous metric perturbation $\delta g_{xx}$ at $\cal{O}$$(\epsilon^2)$. But, because it is inhomogeneous, it feeds back into the resistivity only through interactions with the lattice, which introduces further powers of $\epsilon$ and so can be neglected.}.

\para
 The upshot is that we have three perturbations: $\delta A_x$, $\delta g_{tx}$ and $\delta \phi$, together with the constraint equation that arises from the gauge fixing condition $\delta g_{rx}=0$. Before we jump into a morass of coupled equations, let us first explain some of the physics that underlies these perturbations.
 
\para
We start with the new ingredient which is the scalar perturbation $\delta \phi$. A simple parity argument ensures that the scalar perturbation takes the form,
\be \delta\phi(r,x,t) = \delta \phi(r,t)\sin( k_L x)\nn\ee
However, there is deeper interpretation of this functional form: it is a bulk phonon mode. This is easily seen by rewriting the perturbation as a position dependent phase of the bulk lattice
\be \phi(r,x,t) = \epsilon \phi_0(r)\cos\left(k_L[x-\pi(r,t)]\right) \nn\ee
The phonon mode $\pi$ is related to the scalar perturbation by  $\delta \phi = \epsilon k_L\phi_0(r)\pi(r,t)$.

\para
We require that $\delta \phi(r,t) \sim r^{\Delta_{+}}$ near the boundary, corresponding to a change in the response,  $\delta \langle \cal{O} \rangle$, in the boundary theory. Because we have broken translational invariance explicitly in the boundary theory, there is no Goldstone mode. Instead the response $\delta \langle {\cal O}\rangle$ describes the ``unparticle" soup oscillating in the background of the lattice. It is analogous to cold atoms oscillating in a background optical lattice.

\para In contrast, in the bulk, the phonon $\pi$ is a propagating Goldstone mode. At each radial slice in the bulk, you can think of a layer of material with ``ions" (i.e. peaks of the lattice) positioned at $x - \pi(r,t)= 2\pi n/k_L$. A non-zero momentum in the bulk, $\delta g_{tx}$, collides with these layers and shifts them relative to one another. This disturbance then propagates as a transverse phonon in the radial direction until it  reaches the horizon where the momentum is lost to the system. This simple picture makes it clear that the phonon is responsible for the momentum dissipation in the boundary theory and that this dissipation is ultimately governed by the properties of the horizon. This will be manifest in our formula below for the DC conductivity.

\para
The existence of this bulk phonon mode is intimately tied with the fact that the lattice induces a mass for the graviton. To see this, we can use diffeomorphism invariance to freeze the phonon mode at the expense of introducing a new, propagating degree of freedom in the metric. All we need to is to switch to a new coordinate defined by $\tilde x = x - \pi(r,t)$. This coordinate transformation places the dynamics back into the metric. In this new gauge, $\delta g_{rx}$ becomes dynamical and corresponds to the extra polarisation of a massive graviton. This is entirely analogous to the Higgs mechanism in gauge theory where a would-be Goldstone mode is eaten by the gauge field. Here, instead, the phonon is eaten by the metric. The whole discussion parallels the usual St\"uckelberg formulation of massive gravity \cite{nima}, now with the phonon playing the role of the St\"uckelberg field. (See also \cite{pearson}). 

\para
To truly see that our lattice describes a massive graviton, we should look at the full perturbation equations below. But 
there is a quick, cheap way to get the basic idea. From the discussion above, it is clear that the mass should arise 
from the breaking of translational invariance. In other words, it  comes from the $(\partial_x \phi)^2$ terms in the 
action. Evaluated on the background solution $\phi = \epsilon\,\phi_0(r)\cos(k_Lx)$, the homogeneous contribution to the mass is
\be S_{\rm eff} =  \frac{1}{2} \int d^4x \sqrt{-g} M^2(r)\,g^{xx}\label{masseff}\ee
where the effective mass $M(r)$ is radially dependent and given by
\be M^2(r) = \frac{1}{2}\epsilon^2 k_L^2\phi_0(r)^2\label{mphi}\ee
Expanding out the determinant $\sqrt{-g}$ in \eqn{masseff} will give the promised effective mass to $\delta g_{tx}$ and  $\delta g_{rx}$.
The mass term \eqn{masseff} has the same form as those that arise in the holographic massive gravity theory of \cite{vegh}, albeit with a different radial profile \eqn{mphi}. 


\para
With these basic explanations of the relevant physics in place, let's now turn to the details. As described above, we focus on the homogenous perturbations to leading order in $\epsilon$. To avoid clutter, we'll set $2\kappa^2=L^2=e^2=1$ in what follows. It's simplest to keep the phonon as a physical degree of freedom and  work in $\delta g_{rx}=0$ gauge. The three equations governing the perturbations at order ${\cal O}(\epsilon^2)$ are the Maxwell equation,
\be \left(f \delta A_x'\right)' + \frac{\omega^2}{f}\delta A_x = \frac{\mu}{r_h} \left(r^2\delta g_{tx}\right)'\nn\ee
The scalar equation,
\be r^2\left(\frac{fM^2}{r^2}\pi'\right)' +\frac{\omega^2 M^2}{f}\pi = i\omega\frac{r^2M^2}{f}\delta g_{tx}\nn\ee
and the $r$-$x$ component of the Einstein equation,
\be \left(r^2 \delta g_{tx}\right)' = \frac{\mu r^2}{r_h} \delta A_x +\frac{fM^2}{i\omega} \pi'\nn\ee
There is a further $t$-$x$ component of the Einstein equations but, as usual, the constraints of general relativity mean that this is implied by the three equations above\footnote{In order to satisfy  the full Einstein equations, it is necessary to take into account the homogeneous back reaction of lattice, $A_\mu^{H}$ and  $g_{\mu\nu}^{H}$.}. 

\para The UV boundary condition for the phonon field $\pi$ plays an important role. The fact the we have explicitly, as opposed to spontaneously, broken translational invariance means that we require the fall-off $\pi(r,t) \sim r^{\Delta_+ - \Delta_{-}}$ at the boundary. In contrast, in situations where translational symmetry is broken spontaneously,  the correct boundary condition is that the phonon approaches a constant at the boundary.
\para
It is simple to eliminate $\delta g_{tx}$, leaving two coupled equations for $\delta A_x$ and $\pi$,
\be
(f \delta A_x')' +\frac{\omega^2}{f} \delta A_x  &=&   \frac{\mu^2 r^2}{r_h^2} \delta A_x + \frac{\mu fM^2}{i\omega r_h}\pi' \label{pert1} \\
\frac{1}{r^2}\left(\frac{r^2 f}{M^2}\left( \frac{fM^2}{r^2}\pi'  \right)'\,\right)' +
\frac{\omega^2}{r^2} \pi' &=&  \frac{i\omega \mu}{r_h} \delta A_x + \frac{fM^2}{r^2} \pi'\label{pert2}\ee
The key observation is that these perturbation equations are equivalent to those that arise in massive gravity \cite{vegh,davison,us}\footnote{These equations can be compared, for example, to equations (3.8) and (3.9) of \cite{us}.} with an effective graviton mass $M^2(r)$. The phonon mode $\pi$ is related to the extra propagating metric mode $g_{rx}$ in massive gravity through the relation $\pi' \rightarrow  r^2g_{rx}$. Of course this is not a surprise --- as we have already emphasised, the two descriptions are gauge equivalent.

\section{Conductivity}

To compute the optical conductivity, we need only solve \eqn{pert1} and \eqn{pert2} subject to the appropriate boundary conditions. Fortunately, many of the relevant calculations have already been performed in the context of massive gravity.

\para
The full optical conductivity $\sigma(\omega)$ depends on details of the bulk geometry and gauge field. At small lattice strength it exhibits a Drude peak, as  plotted\footnote{The plots were made with  $m^2L^2=-0.25 $, $k_L=0.433 \mu$ and $\epsilon =0.1 $. Strictly speaking, we should compute the ${\cal O}(\epsilon^2)$ corrections to the backgrounds before determining the $\omega\neq 0$ conductivity but, by eye, the plots are identical}  numerically in Figure 1.

\DOUBLEFIGURE{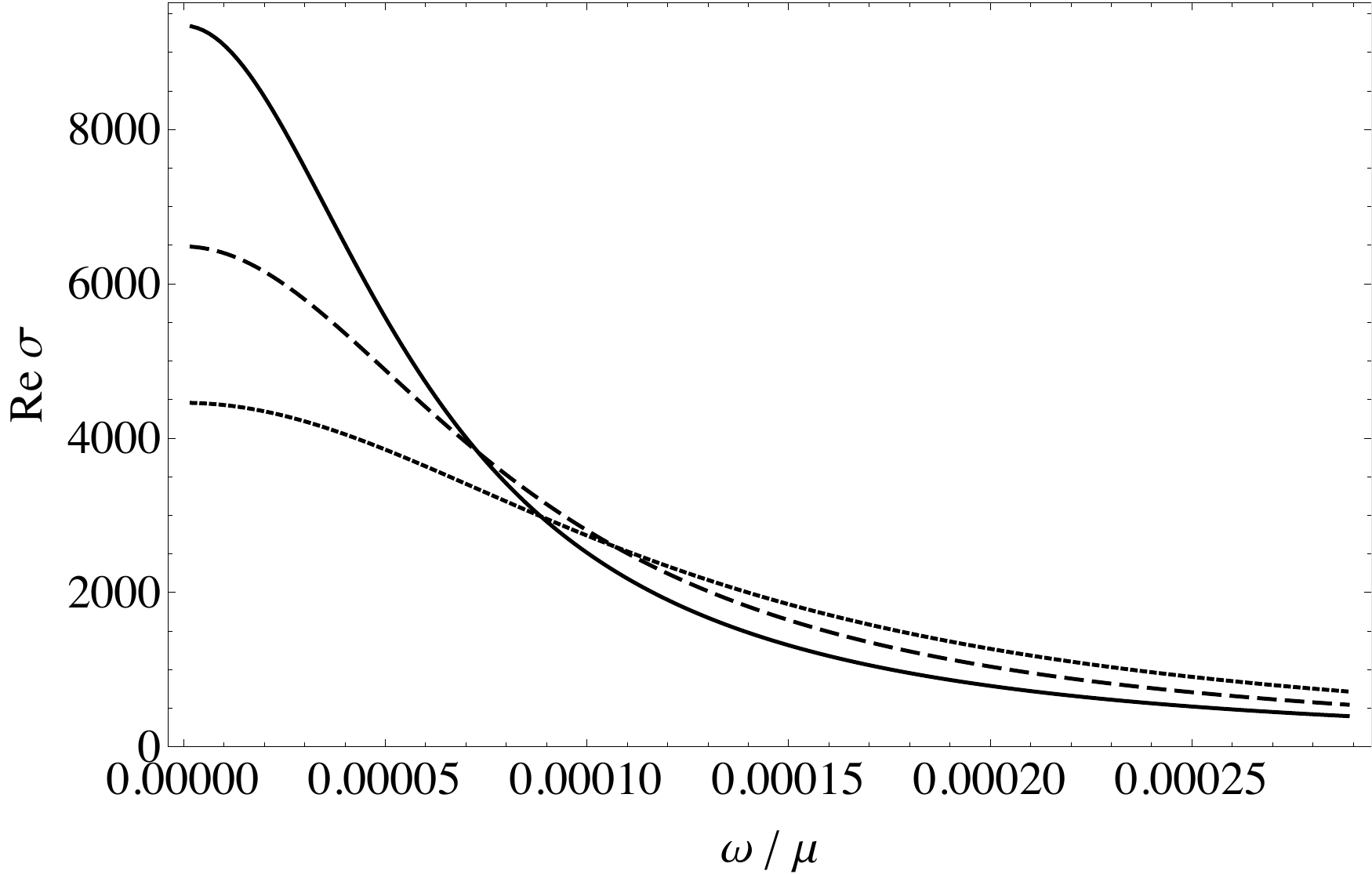,width=200pt}{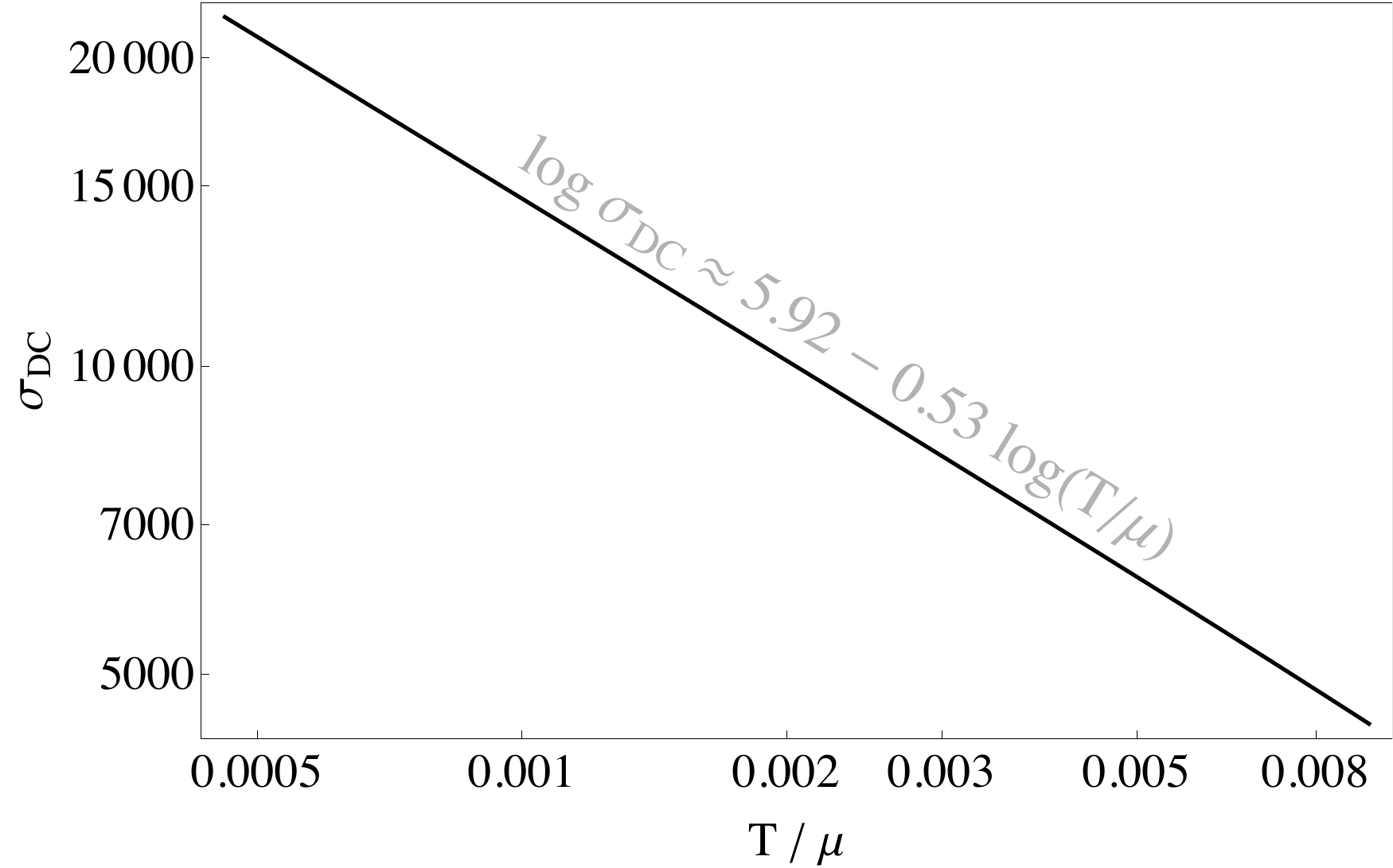,width=200pt}
{The optical conductivity shown, in descending order, for $T/\mu = 0.0023$, $0.0046$ and $0.0092$}{A log-log plot of the  DC conductivity. The analytic prediction for these values is $\sigma_{DC} \sim T^{-0.5275}$.}
\noindent

\para
The DC conductivity is computed numerically in Figure 2. However, here we can make more analytic progress. This is  because, as shown in \cite{us}, the DC conductivity depends only on the behaviour of the fields at the infra-red horizon. The argument is a generalisation of an earlier observation by Iqbal and Liu \cite{il}. The essence of it goes as follows: the photon $\delta A_x$ has an effective mass proportional to the charge density $\mu/r_h$; meanwhile, as we described above, the phonon has a mass proportional to $M^2$. However, the two modes mix. And it is simple to check that there is a linear combination which is massless and, in the $\omega\rightarrow 0$ limit, does not evolve from the horizon to the boundary. Furthermore, this linear combination carries the information about the conductivity. This means that one can compute the DC conductivity in terms of properties of the horizon of the black hole. 

\para
For completeness, we provide a more detailed review of the above calculation in the Appendix. The end result is that the scattering rate is fixed by graviton mass evaluated at the horizon \cite{us},
%
%
\be
\Gamma = \frac{s}{4 \pi} \frac{M^2(r_h)}{\cal{E} + \cal{P}} \label{result}
\ee
where the entropy density $s$, energy density ${\mathcal{E}}$ and pressure $\cal{P}$ are thermodynamic functions that are non-zero in the extremal RN black hole background. This result was also obtained for hydrodynamic transport in massive gravity in \cite{davison}. 
The fact that this scattering rate is already proportional to $\epsilon^2$ through the graviton mass is ultimately why were able to ignore the homogeneous corrections to the RN geometry. These would only affect the thermodynamic factors, and hence the scattering rate, at higher order. 
\para The key content of this formula is that the scattering rate is simply determined by the effective graviton mass induced by lattice:
\be \Gamma \sim M^2(r_h) \sim \epsilon^2 k_L^2 \phi_0(r_h)^2 \nn\ee
where we have dropped the other coefficients on the grounds that they are, to leading order, constants that are independent of temperature.

\para
All that remains is to determine the infra-red behaviour of the scalar profile $\phi_0(r)$ which will govern the temperature dependence of graviton mass \eqn{mphi}.
But this is straightforward. At $T=0$, the infra-red geometry is AdS${}_2\times {\bf R}^2$. As we reviewed in the introduction, this is the holographic manifestation of a locally critical theory. If we denote the radial coordinate in AdS${}_2$ as $\zeta$, the regular solution for $\phi$ falls off asymptotically in the infra-red as
\be \phi_0 \sim \zeta^{\half-\nu_{k_L}}\nn\ee
where  $\nu_{k_L}+1/2$  is the dimension of the dual operator ${\cal O}(k_L)$ in real space, with the dependence on the lattice spacing given by  $\nu_{k_L}= \sqrt{ {1\ov 4} + \frac{1}{6} m^2 L^2 + 2 k^2_L/\mu^2 } $. Upon taking a Fourier transform, the dimension of the operator in frequency space becomes $\Delta_{k_L} = \nu_{k_L}-1/2$, so we have
\be \phi_0 \sim \zeta^{-\Delta_{k_L}}\nn\ee
 At finite temperature, the $AdS_2$ geometry terminates in a horizon at $\zeta_H \sim T^{-1}$. This means that the effective graviton mass, and hence resistivity, scales as
\be
\rho \sim \epsilon^2 k_L^2 T^{2 \Delta_{k_L}}\label{answer}
\ee
Happily, this is precisely the result of Hartnoll and Hofman \cite{sandiego} that we reviewed in the introduction.

\section{Closing Remarks}

\para Throughout this paper, we have relied on the technical crutch of the small-lattice expansion. This allowed us to isolate the phonon mode as the relevant, extra degree of freedom in computing the resistivity.  However, we would like to suggest that, even in more complicated situations, the phonon mode continues to dominate the low-temperature physics. Here we offer some suggestions on how this may happen. 

\para 
Let us first address what would happen if we compute the resistivity to higher order in the lattice strength, $\epsilon$. Further fields ---  including, most pertinently, the gauge field $A_t$ --- will pick up a spatial modulation and therefore contribute to the effective mass of the graviton at ${\cal O}(\epsilon^4)$. The analysis of \cite{sandiego} shows that each such field will give a contribution to the DC conductivity of the form \eqn{sd}. At low temperatures, the charge density $J^t(2k_L)$  is the least irrelevant operator (together with $T_{tt}$, with which it mixes) to get a spatially modulated expectation value and so, although it is sub-leading in the $\epsilon$-expansion, dominates the low-temperature resistivity \cite{sandiego,jorge1}.

\para
Although technically more involved, it seems clear how the field theory expectations above are mirrored in the gravity calculation. Clearly,  we will have many more perturbed fields in the game. However, among these we expect that there remains a linear combination which is massless and, therefore, does not evolve from the horizon to the boundary. This means that we can focus attention on the far infra-red geometry.  Here, the gauge field $A_t$ is the largest spatially-varying field and the fields dominating the perturbation equations are $\delta A_x$, $\delta g_{tx}$ and now the phonon  $\delta A_t$ arising from the induced ionic lattice. Thus, in the far IR, the perturbation equations reduce to those considered here and resistivity will again be given by \eqn{answer}, but with the exponent $\Delta_k$ replaced by the appropriate dimension of the ionic lattice (which was computed in \cite{jottar}).

\para
We note that the conceptual steps sketched above also hold for other situations, such as the ionic lattice, where no simple expansion in the lattice strength is available. Instead, we replace the expansion in $\epsilon$ with an infra-red expansion. Of course, this is what one naturally expects for the DC conductivity and,  even without an explicit demonstration of the massless mode, it should be possible to extract the leading temperature dependence of the resistivity by a matching calculation \cite{ads2,linearresistivity}.  

\para 
Moving beyond the $AdS_2 \times {\bf R}^2$ infra-red geometries, there are other ``hyperscaling violating"  geometries which, while exhibiting local criticality, do not suffer from the pathology of a ground state entropy \cite{kevin,elias}. Rather, the horizon radius scales as some power of temperature $s \sim r_h^{-2} \sim T^{\eta}$.  In the context of massive gravity, the DC conductivity can again be computed exactly \cite{us}, but now there is a temperature dependence even if the mass of the graviton is constant. (See also \cite{leiden}).  It is a simple matter to repeat the calculations above for these geometries. From the scalar wave equation, one finds that the IR behaviour of the lattice  field is now given by 
 $\phi_0(r) \sim r^{1 + 1/\eta - 2 \nu_{k_L}/ \eta}$. From our general result \eqn{result} we can then deduce that the leading temperature dependence of the resistivity is given by
\be
\rho \sim \frac{\phi_0^2(r_h)}{r_h^2} \sim T^ {2 \nu_{k_L} - 1} \nn
\ee 
This is in agreement with the scaling derived using the memory matrix formalism \cite{shaghoulian, martin}.

\para 
Finally, there is one last issue that we would like to address: what happens in situations in which breaking of translational symmetry occurs spontaneously\footnote{We thank Sung-Sik Lee for prompting us to think about this question.}? There are a number of holographic examples of spontaneous lattice formation, including \cite{hirosi,bolog,aristos}. The perturbation equations that we derived above continue to hold, with one important difference: the UV boundary condition for the phonon becomes $\pi\sim {\rm const.}$ rather than the fall-off  $\pi \sim r^{\Delta_{+} - \Delta_{-}}$ required for explicit breaking. With these boundary conditions, the derivation of the DC conductivity presented in the Appendix no longer holds. Indeed, solving the perturbation equations numerically with this new boundary condition, we find that the delta function in the conductivity is restored. This is the expected behaviour in models with spontaneously broken translational invariance. 

\section*{Acknowledgements}

Our thanks to Richard Davison, Sean Hartnoll, Sung-Sik Lee, Don Marolf, Koenraad Schalm and Jan Zaanen for useful conversations. We are grateful to the Newton Institute for hospitality while this project was undertaken.
MB and DT are supported by STFC and by the European
Research Council under the European Union's Seventh Framework Programme
(FP7/2007-2013), ERC Grant agreement STG 279943, ``Strongly Coupled Systems''.

\appendix
\section{Appendix}

In this Appendix, we provide a derivation of the result \eqn{result}. The derivation presented here is simpler  than that originally given in \cite{us}, but at the cost of being slightly less rigorous. (Specifically, in a number of places we will assume that the DC conductivity is finite; readers that find this unsatisfactory can return to the original proof of  \cite{us}).

\para
To make contact with the result of \cite{us}, we change gauge and work with the metric component $\delta \tilde{g}_{rx} = f(r) \delta g_{rx} = f(r) \pi'/r^2$. 
The equations of motion \eqn{pert1} and \eqn{pert2} can then be written in the schematic form
\be
\left(\begin{array}{cc} L_1 & 0 \\ 0 & L_2\end{array}\right)
\left(\begin{array}{c} \delta A_x \\ \delta \tilde{g}_{rx}\end{array}\right) +
\frac{\omega^2}{f}\left(\begin{array}{c} \delta A_x \\
\delta \tilde{g}_{rx}\end{array}\right)= {\cal M}\left(\begin{array}{c} \delta A_x \\
\delta \tilde{g}_{rx}\end{array}\right)\label{matrixpert}
\ee
where $L_i$ are linear differential operators and ${\cal M}$ is a mass matrix whose exact form can be found in \cite{us}. The universality of the DC conductivity hinges on the observation that ${\rm det}\,{\cal M}=0$. This means that there exists a massless eigenmode of the differential equations that is some linear combination of  $\delta A_x$ and $\delta \tilde{g}_{rx}$. The equation of motion for this mode, which appears above  Equation (3.12) in \cite{us}, can be written in the form 
\be
 \nn
   \Pi' + {\omega^2 \ov f} \le(  \delta A_x - \frac{\mu r^2}{i\omega r_h}\delta \tilde{g}_{rx} \ri) = 0
\ee
In the $\omega\rightarrow 0$ limit, this allows us to deduce that there is a conserved quantity $\Pi$ 
\be
  \Pi(r) = f(r)\le[ \delta A_x' - \frac{\mu r^2}{i \omega r_h M^2} \le( M^2  \delta \tilde{g}_{rx} \ri)' \ri] \label{pi}
\ee
When translational symmetry is broken explicitly by a source, the boundary condition on the phonon translates into the condition that $\le( M^2  \delta \tilde{g}_{rx} \ri)' = 0$ at the boundary. This, in turn,  implies that for all scalar fields above the BF bound, the second term in \eqn{pi} is subleading in the UV.  We can therefore identify the boundary value of $\Pi$ with the current in the boundary theory, that is $\Pi(r=0) = \delta A_x'(r=0)$. 
This allow us to use $\Pi$ to define a membrane conductivity associated with each radial slice via
\be \sigma_{DC}(r) = \lim_{\omega\rightarrow 0} \frac{\Pi}{i \omega \delta A_x} \label{sigma}
\nn
\ee
which reduces to the conductivity of the boundary theory as $r\rightarrow 0$.

\para 
The next step of the argument is to show that $\sigma_{DC}(r)$ is independent of $r$. We have already seen that, at low frequencies, $\Pi$ is a constant in the bulk. Under the assumption that the DC conductivity is finite, we must have  $\delta A_x \sim {\cal O}(1)$ and  $\delta A_x' \sim {\cal O}(\omega)$ in the bulk. We can therefore also take $\delta A_x$ to be a constant at leading order, and hence the DC conductivity $\sigma_{DC}(r)$ is indeed independent of the radial position. 

\para 
All that remains is to evaluate $ \sigma_{\rm DC}(r)$ near the horizon. Ingoing boundary conditions mean that the gauge perturbation oscillates as $ \delta A_x(r)  = f(r)^{-{i\omega/4 \pi T}}$. The behaviour of $\delta \tilde{g}_{rx}$ can be deduced from \eqn{pert1}. Once again, assuming a finite DC conductivity, the two terms on the right-hand side of \eqn{pert1} must cancel to leading order in $\omega$ to allow $\delta A_x'\sim {\cal O}(\omega)$.  This means that
\be
   \delta \tilde g_{rx}(r) =  -\frac{i\omega\mu}{r_h M^2(r)} \, \delta A_x(r) + {\cal O}(\omega^2)\nn
\ee
With these two results, it is a simple matter to evaluate $\sigma_{DC}$ at the horizon. It is given by
\be
\sigma_{DC} =  \bigg( 1 + \frac{\mu^2}{M^2(r_h)} \bigg) \nn
\ee
For the lattice models of interest in the current paper, we can only trust this computation of the resistivity to order $\epsilon^2$. This means that, upon inverting, we drop the $+1$ above.
From the conductivity, we can extract the scattering time  via the identification \cite{nernst}
\be
\sigma_{DC} = \frac{Q^2}{\cal{E} + \cal{P}} \frac{1}{\Gamma}\nn
\ee
where ${\cal Q}$ is the charge density. For our holographic models, the charge density is related to the chemical potential by $\mu^2 = {\cal Q}^2r_h^2= 4 \pi {\cal Q}^2 /s $ where $s$ is the entropy density. This then  gives the result \eqn{result}.

\end{document}